\documentclass[twocolumn,showpacs,preprintnumbers,amsmath,amssymb]{revtex4}

\usepackage{graphicx}
\usepackage{dcolumn}
\usepackage{bm}

\begin{document}

\title{Photoproduction of Quarkonium in Proton-Proton and
Nucleus-Nucleus Collisions}

\author{Spencer R. Klein}
\affiliation{%
Lawrence Berkeley National Laboratory, \\
Berkeley, CA 94720, USA
}

\author{Joakim Nystrand}
\affiliation{
{Department of Physics, University of Bergen} \\
N-5007 Bergen, Norway 
}

\begin{abstract}

We discuss the photoproduction of $\Upsilon$ and $J/\psi$ at high
energy $\overline{p}p$, $pp$ and heavy ion colliders.  We predict
large rates in $\overline{p}p$ interactions at the Fermilab Tevatron
and in $pp$ and heavy-ion interactions at the CERN LHC.  The $J/\psi$
is also produced copiously at RHIC.  These reactions can be used to
study the gluon distribution in protons and heavy nuclei.  We also
show that the different $CP$ symmetries of the initial states lead to
large differences in the transverse momentum spectra of mesons
produced in $\overline{p}p$ vs. $pp$ collisions.

\end{abstract}


\maketitle

Photoproduction has traditionally been studied with fixed target
photon beams, at electron-proton colliders, and, to a limited extent, 
at relativistic heavy ion colliders.  However, energetic protons also
have large electromagnetic fields, and high energy
$pp$ and $\overline{p}p$ colliders can be used to study
photoproduction, at photon energies higher than are currently
accessible.  These photoproduction reactions are of interest as a way
to measure the gluon distribution in protons at low Feynman$-x$.

In this letter, we study photoproduction of heavy quark vector mesons
in $\overline{p}p$ collisions at the Fermilab Tevatron and in $pp$
collisions at the Relativistic Heavy Ion Collider at Brookhaven and at
the Large Hadron Collider (LHC) at CERN.  At the Tevatron, 
the high rates allow for detailed measurements of gluon distributions 
around $x \approx 1.5 - 5 \times 10^{-3}$, and at the LHC $x \approx 2 - 7
\times 10^{-4}$ can be reached. We also discuss the coherent
photonuclear production of $\Upsilon$ in nucleus-nucleus collisions at
RHIC and the LHC.

We show that the different $CP$ symmetry between the $pp$ and 
$\overline{p}p$ initial states leads to large differences in 
the transverse momentum, $p_T$ spectra of the produced mesons.
Finally, we discuss how these events can be separated 
from purely hadronic interactions. 

We use data from HERA and fixed target experiments on exclusive
photoproduction of heavy vector mesons in photon-proton
interactions \cite{Crittenden:1997yz} as input to our calculations.
Because data on the $\Upsilon$ is limited, we use QCD based models as
a basis for parameterizations of the cross sections.  The paucity of
experimental data on photoproduction of the $\Upsilon$ leads to a
relatively large uncertainty in the parameterized cross sections, but
is also a strong motivation for investigating new production channels.

The total $J/ \Psi$ photon-proton cross section for quasi-real ($Q^2
\approx 0$) photons has been measured from near threshold up to
photon-proton center-of-mass energies, $W_{\gamma p}$, 
up to 200 GeV. The cross section increases with $W_{\gamma p}$ roughly as
$W_{\gamma p}^{0.8}$.  We parameterize $\sigma_{\gamma p}(W_{\gamma
p}) = 1.5 \cdot W^{0.8}$ [nb] (W in GeV). A drawback of this
parameterization is that there is a discontinuity at the threshold
energy, $W_{\gamma p}=m_p + m_{J/ \Psi}$.

The two measurements of the $\Upsilon$ at HERA both have significant
uncertainties.  The Zeus collaboration measured $\sigma \cdot Br(
\Upsilon \rightarrow \mu \mu ) = 13.3 \pm 6.0 $(stat.)$ ^{+2.7} _{-2.3}
$(syst.)~pb at a mean center-of-mass energy of $\langle W_{\gamma
p}\rangle = $120 GeV \cite{Breitweg:1998ki}.  The H1 collaboration
found $\sigma \cdot Br(\Upsilon \rightarrow \mu \mu) = 19.2 \pm 9.9
$(stat.)$ \pm4.8 $(syst.)~pb at $\langle W_{\gamma p}\rangle= $143
GeV \cite{Adloff:2000vm}.  Both experiments estimate that roughly 70\%
of the signal comes from the $\Upsilon(1S)$ state.

The leading-order expression for the photoproduction of a vector meson
of mass $M_V$ is \cite{Ryskin:1992ui}
\begin{equation}
\left. \frac{d \sigma (\gamma p \rightarrow Vp)}{dt} \right|_{t=0} =
\frac{\alpha_s ^2 \Gamma_{ee}}{3 \alpha M_V ^5} 16 \pi^3 \left[ x
g(x,M_V^2/4) \right]^2.
\label{Ryskin}
\end{equation}
Two more sophisticated calculations have considered the use of
relativistic wave functions, off-diagonal parton distributions, and
NLO contributions \cite{Frankfurt:1998yf,Martin:1999rn}.  Although the
approaches differ, the final results are in good agreement.  The
cross section for $\Upsilon$(1S) production scales 
roughly as $W_{\gamma p}^{1.7}$.  We use a parameterization which is
consistent with both HERA results: $\sigma_{\gamma p}(W_{\gamma p}) =
0.06 \cdot W_{\gamma p}^{1.7}$ [pb] (W in GeV).  An alternative method
based on parton-hadron duality gives cross sections $\sim$30-50\%
larger depending on $W_{\gamma p}$ \cite{Martin:1999rn}. 
Our calculations are for the $\Upsilon$(1S). 

We estimate the uncertainties in the $\Upsilon$ cross section by
fitting the H1 and Zeus data to 
the function $\sigma_{\gamma
p}(W_{\gamma p}) = C \cdot W_{\gamma p}^{1.7}$.  The constant $C$ is
determined from two 
fits, one with the experimental errors
(quadratic sum of statistical and systematical) added to the measured
value and 
the other with the experimental errors subtracted. These fits give
$C=0.175$ [pb] and $C=0.054$ [pb], respectively.

The cross section to produce a vector meson in a proton-proton
collision is 
\begin{equation}
   \sigma(p+p \rightarrow p+p+V) = 2 \int_{0}^{\infty} \frac{dn}{dk}
   \, \sigma_{\gamma p} (k) \, d k \; .
\label{eq:sigma}
\end{equation}
As will be discussed below, quantum mechanical interference
alters the transverse momentum ($p_T$) spectrum, but does not affect
the total cross section significantly.

We use the photon spectrum from Ref. \cite{Drees:1988pp}: 
\begin{eqnarray}
\frac{dn}{dk} = && \frac{\alpha}{2 \pi k} \left[ 1 + (1 - 
\frac{2k}{\sqrt{s}})^2 \right] 
\left( \ln{A} - \frac{11}{6} + \frac{3}{A} \right. \nonumber\\
&& \left. - \frac{3}{2 A^2} + \frac{1}{3 A^3} \right) 
\label{eq:pdndk}
\end{eqnarray}
where $A = 1 + (0.71 GeV^2)/Q_{min}^2$ and $Q_{min}^2 \approx (k/
\gamma)^2$. It is derived using a proton form factor, $F(Q^2) = 1/(1 +
Q^2/(0.71 GeV^2) )^2$. This spectrum is close to that of a point
charge with a minimum impact parameter of $b_{min} = 0.7$ fm.  It
corresponds to emission of a photon with the proton remaining
intact. For exclusive vector meson production, the target proton must
also remain intact.  This stricter requirement slightly decreases 
the effective photon flux. To estimate this
uncertainty, we have also performed calculations with a photon
spectrum corresponding to a point charge with a minimum impact
parameter of $b_{min} = 1.0$ fm.

\begin{figure}
\includegraphics[width=8cm]{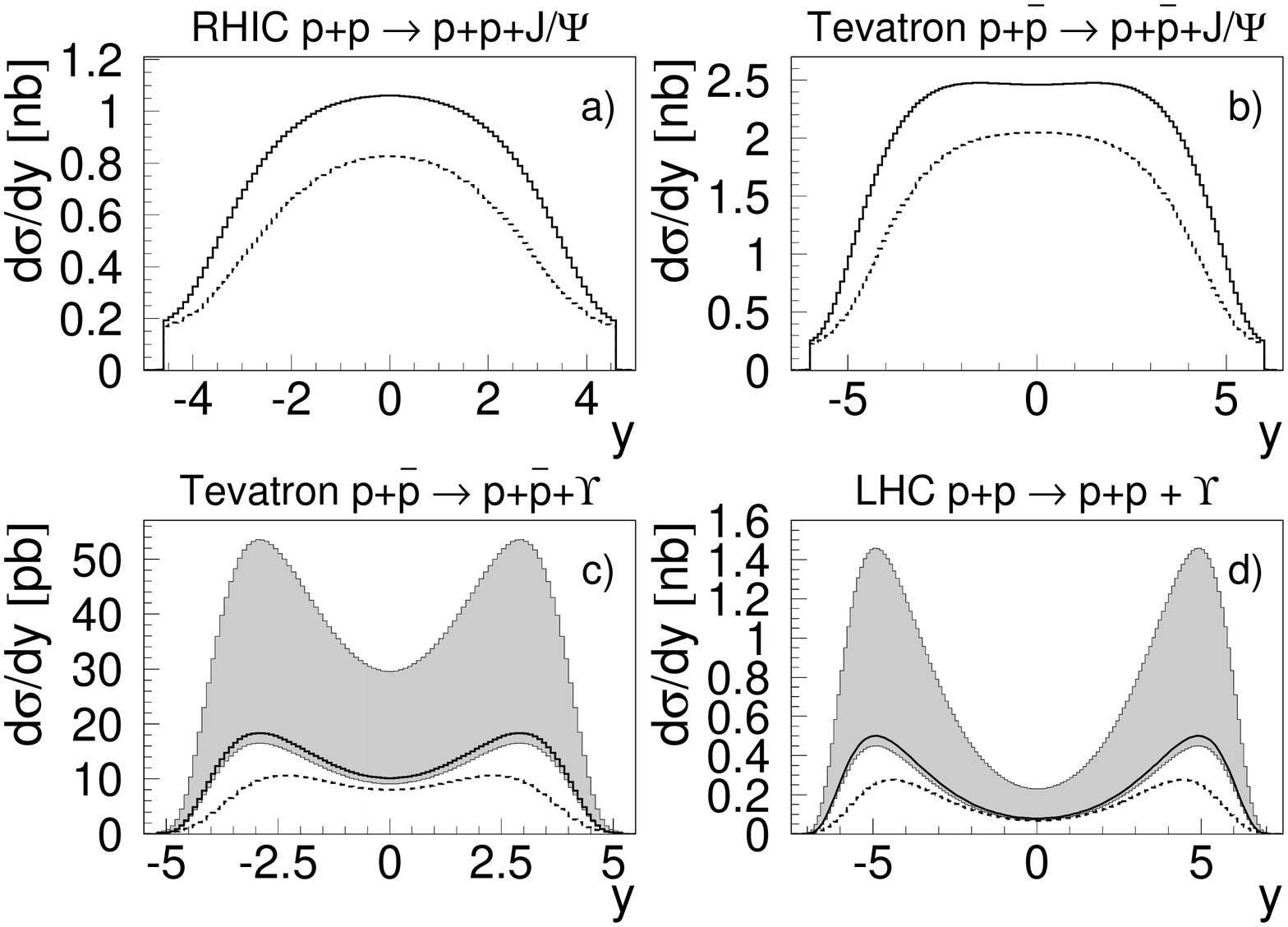} 
\caption{\label{fig:pp} Rapidity distributions for photoproduction of
$J/ \Psi$ and $\Upsilon$(1S) mesons in $p p$ and $p \overline{p}$
interactions at RHIC, the Tevatron, and the LHC. The curves are
explained in the text.}
\end{figure}

The rapidity, $y$, of a produced state with mass $M_V$ is related to
the photon energy through $y = \ln(2k/M_V)$. Using this relation in
Eq.~\ref{eq:sigma} and differentiating gives
\begin{equation}
\frac{d \sigma}{dy} = k \frac{dn}{dk} \sigma_{\gamma A \rightarrow V A} (k) \; .
\label{eq:rapidity}
\end{equation}
Interchanging the photon emitter and target corresponds to a
reflection around $y=0$; the total cross section is the sum of the two
possibilities.  The rapidity distributions are shown in
Fig.~\ref{fig:pp}.  The calculations are for collision energies of
$\sqrt{s} = 500$ GeV at RHIC, $\sqrt{s} = 1.96$ TeV at the Tevatron,
and $\sqrt{s} = 14$ TeV at the LHC. The solid and dashed histograms
are for the parameterizations $\sigma_{\gamma p}(W_{\gamma p}) = 1.5
\cdot W_{\gamma p}^{0.8}$ [nb] and $\sigma_{\gamma p}(W_{\gamma p}) =
0.06 \cdot W_{\gamma p}^{1.7}$ [pb] for the $J/ \Psi$ and $\Upsilon$,
respectively.  The solid histogram is for the photon spectrum in
Eq. \ref{eq:pdndk}, while the dashed histogram is for a point charge
with a cut-off at $b_{min}=1.0$~fm.  The grey band shown for the
$\Upsilon$ corresponds to the two fits to the data described above
(with the photon spectrum of Eq. \ref{eq:pdndk}).  The sharp cut-off
at large rapidities for the $J/\Psi$ is due to the discontinuity at
threshold in the parameterization of $\sigma_{\gamma p}$. Using the
photon spectrum in Eq. \ref{eq:pdndk}, the total cross sections for
the $\Upsilon$ are 12 pb, 120 pb, and 3.5 nb at RHIC, the Tevatron,
and the LHC, respectively.  For the $J/\psi$, the corresponding cross
sections are 7.0 nb, 23 nb, and 120 nb.

Exclusive $J/\psi$ production in $pp$ interactions was previously
considered by Khoze {\it et al.}\cite{Khoze:2002dc}. They use a very
different approach, based on the proton energy lost.  Analytical or
numerical comparisons between their results and ours are therefore
difficult.

At the design luminosity for $p \overline{p}$ at the Tevatron ($2
\times 10^{32} cm^{-2} s^{-1}$) \cite{PDG}, the $\Upsilon$ production
rate is 76/hr. Even with the 2.5\% per flavor branching ratios to
$l^+l^-$, good signals should be observable at the Tevatron.  For
example, in 1 year (8,000 hr) at 33\% overall efficiency,
5,000 decays should be observable in each of the $e^+e^-$ and
$\mu^+\mu^-$ channels.  At RHIC, the lower energy, luminosity and
shorter $pp$ running time decrease the signal. However, with the
planned RHIC II luminosity upgrade, $\Upsilon$ photoproduction could
be studied. The situation is better for the $J/ \Psi$, where the
production rate at RHIC is 250/hr.  At the LHC, the $J/ \Psi$ and 
$\Upsilon$ rates are both high.

The mid-rapidity photoproduction cross section for $\Upsilon$ at the
Tevatron is about 0.1\% of the hadronic inclusive $\Upsilon$ cross section \cite{Abe:1995an}.
Similarly, the photoproduction cross section at mid-rapidity for the
$J/\Psi$ at RHIC is about 0.1\% of the hadronic inclusive cross
section measured at $\sqrt{s} = 200$ GeV (the total cross section is
expected to be about twice as large at $\sqrt{s} = 500$ GeV)
\cite{Adler:2003qs}.

Although the photoproduction cross section is a small fraction of the
hadronic cross section, separation of this reaction channel seems
possible given the very different character of the photon induced events. 
We will discuss some of the selection criteria, and estimate their
effectiveness at rejecting hadronic events.

Hadronically produced vector mesons have $p_T\sim M_V$.  In contrast,
almost all of the photoproduced mesons have $p_T < 1$ GeV/c
(cf. Fig. 2).  A $p_T < 1$ GeV/c cut eliminates about 94\% of the
hadroproduced $\Upsilon$ at the Tevatron \cite{Abe:1995an}, while
retaining almost all of the photoproduction.

As long as both protons remain intact, the vector meson will not be
accompanied by any other particles in the same event. In contrast, in
hadronic events, the produced particles are distributed over the
available phase space.  If the average charged particle multiplicity
is $\langle dn_{ch}/dy\rangle$, then the probability of having a
charged particle free rapidity gap with width $\Delta y$ is $\exp(-
\Delta y \cdot \langle dn_{ch}/dy\rangle)$. The mean particle
densities at mid-rapidity are $\langle dn_{ch}/dy\rangle = 3.0$ at
$\sqrt{s}=500$ GeV, $\langle dn_{ch}/dy\rangle = 4.0$ at $\sqrt{s}=
1.96$ TeV, and $\langle dn_{ch}/dy\rangle \approx 5.5$ at $\sqrt{s}=
14$ TeV (extrapolated), neglecting any possible difference between $p$
and $\overline{p}$ \cite{Abe:1989td}.  Requiring that the vector meson
be surrounded by particle free regions (rapidity gaps) with a total
width $\Delta y = 3.0$ will reduce the background by a factor of
$\approx 10^{-4}$ at RHIC, $\approx 10^{-5}$ at the Tevatron, and
$\approx 10^{-7}$ at the LHC.  The total width can be split between
two or more gaps, e.g. two gaps of width 1.5, or one of width 3.0.
These gaps fit within the acceptance of existing and planned
detectors, and provide more rejection power than is needed.  The
selection could be improved by using calorimetry to detect neutral
particles in the gaps.

The CDF collaboration has identified a sample of exclusive $J/\psi$s
\cite{angela}; they do not give a cross section, but most of the yield
is in the region $p_T< 1$~GeV/c, as expected for photoproduction.

Exclusive vector meson production in $p p$ and $p \overline{p}$
collisions differs from production in $ep$ or $eA$ collisions in that
both projectiles can act either as target or photon emitter.  For very
small momenta of the produced state, one cannot distinguish which
proton (or anti-proton) emitted the photon and which acted as target,
so adding the cross sections is not justified. This interference was
studied for nucleus-nucleus collisions \cite{Klein:1999gv}.

The interference is best understood in a plane perpendicular to the
direction of the beams.  The impact parameters for photoproduction are
typically a few fm because of the long range of the electromagnetic
force, but the production is always localized to one of the two
projectiles.  Because of symmetry, the effect is largest for $y=0$.

\begin{figure}
\includegraphics[width=8cm]{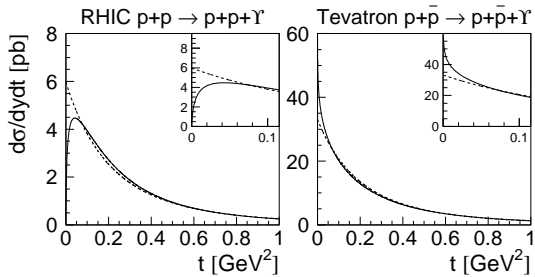}
\caption{\label{fig:interference} $d \sigma / dy dt$ for photoproduced
$\Upsilon$(1S) mesons at mid-rapidity in $p p$ and $p \overline{p}$
collisions at RHIC and the Tevatron. The inset has an expanded $t$
scale.}
\end{figure}

The differential cross section, $d \sigma / dy dt$, where $t$ is the
momentum transfer from the target ($t \approx p_T^2$), 
may be written as an integral over the impact parameter, $b$:
\begin{equation}
   \label{interference}
   \frac{d \sigma}{dy dt} = \int_{b>b_{min}} \left| A_1  +  A_2 \right|^2  d^2 \vec{b} \; .
\end{equation}
$A_1$ and $A_2$ are the amplitudes for production off each of the two
targets.  At mid-rapidity, considered here,
$|A_1| = |A_2|$. The amplitude is normalized to the cross section for
a single source (Eq.~\ref{eq:rapidity}).

If the produced vector mesons are treated as plane waves, $A_i = A_0
\exp( i \vec{p} \cdot \vec{x} )$, the total amplitude is
\begin{equation}
\label{eq:costerm}
\left| A_1 + A_2 \right|^2 = 2 |A_1|^2 \left( 1 \pm \cos( \vec{p}_T \cdot \vec{b} ) \right) \; .
\end{equation}
The sign of the cos-term depends on the symmetry of the system.  In a
$pp$ collision, moving the vector meson emission (scattering) from one
proton to the other corresponds to a parity transform. 
In a $p \overline{p}$ collision, however, it corresponds to a
charge-parity (CP) operation. Since the vector meson has quantum
numbers $J^{PC} = 1^{--}$, the interference is destructive for $pp$
and constructive for $\overline p p$. The sign in Eq.~\ref{eq:costerm}
is ``$-$'' in $p p$ collisions (as in nucleus-nucleus
collisions \cite{Klein:1999gv}) and ``$+$'' in $p \overline{p}$
collisions. With adequate statistics, the interference might be used
to search for $CP$ violation.

This interference alters the vector meson $p_T$ spectrum 
near mid-rapidity. Without interference, the $p_T$ spectrum is that
for production off a single (anti-)proton. This spectrum is the
convolution of the photon transverse momentum spectrum with the
spectrum of transverse momentum transfers from the
target \cite{Klein:1999gv}.  For $p_T > \hbar/\langle b\rangle$, the 
$\cos(\vec{p}_T \cdot \vec{b} )$-term in Eq.~\ref{eq:costerm} oscillates
rapidly as $b$ varies, and the net contribution to the integral will
be zero. For small transverse momenta, however, $p_T \ll 1 / \langle b
\rangle$, $\vec{p} \cdot \vec{b} \approx 0$ for all relevant impact
parameters, and interference alters the spectrum. This is illustrated in
Fig.~\ref{fig:interference}, which compares $d \sigma / dy dt$ with and 
without interference at RHIC and the Tevatron. For
Fig.~\ref{fig:interference}, we imposed a cut $b_{min}=1.0$ fm; this
has a small effect on the spectrum.
The interference is large for $t < 0.05$~GeV$^{2}$/$c^{2}$. The different 
sign of the interference in $p p$ and $p \overline{p}$ is clearly visible. 

In addition to $pp$ and $p \overline{p}$ interactions, vector mesons
are produced in coherent ultra-peripheral nucleus-nucleus
collisions \cite{Klein:1999qj}. The STAR collaboration has 
observed $Au+Au \rightarrow Au+Au+\rho^{0}$ at RHIC \cite{Adler:2002sc}.  
With a cut on $p_T<100 MeV/c$, the signals were quite clean.

For coherent production, the momentum transfer from the nucleus to the
vector meson is determined by the nuclear form-factor. 
Since the $\Upsilon(1S)$ has a small cross section to interact with
a nucleon, hadronic shadowing should be negligible for it.
The forward scattering amplitude scales with the
number of nucleons, $A$, squared:
\begin{equation}
\left. \frac{d \sigma (\gamma A \rightarrow \Upsilon A)}{dt} \right|_{t=0} = A^2 \, 
\left. \frac{d \sigma (\gamma p \rightarrow \Upsilon p)}{dt} \right|_{t=0} \, | F(t) |^2 \; .
\label{A2_scaling}
\end{equation}
A Woods-Saxon distribution is used for the nuclear form factor $F(t)$.
The total photonuclear cross section is the integral of
Eq. \ref{A2_scaling} over all momentum transfers, $t > t_{min} =
[M_{\Upsilon}^2/4 k \gamma]^2$. The input to the calculation is again
the parameterizations of the total photon-proton $\Upsilon$ cross
section discussed above.  When determining the forward scattering
amplitude from the total photon-proton cross section, an exponential
t-dependence is assumed with the same slope, 4 GeV$^{-2}$, as for $J/
\Psi$ production.  This leads to a forward scattering amplitude about
5\% lower than if the proton form factor above had been used.

\begin{figure}
\includegraphics[width=8cm]{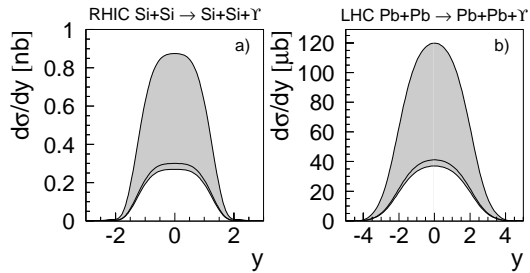} 
\caption{\label{fig:AA} Rapidity distributions of $\Upsilon$ mesons
produced in coherent photonuclear interactions at RHIC and the
LHC. The solid curves correspond to the parameterization
$\sigma_{\gamma p} (W_{\gamma p}) = 0.06 \cdot W^{1.7}$ [pb], and the
grey bands show the uncertainty in $\sigma_{\gamma p}$.
}
\end{figure}

With this, and the photon spectrum in \cite{Klein:1999qj}, the total
cross section and the rapidity distributions can be calculated. 
Figure~\ref{fig:AA} shows $d\sigma/dy$ for $\Upsilon$ production in
Si+Si interactions at RHIC ($\gamma = 135$) and Pb+Pb interactions at
the LHC ($\gamma = 2940$). The total cross section (solid curve)
is 0.72 nb for Si+Si at RHIC and 170
$\mu$b for Pb+Pb at the LHC.

Because of the low photon flux,
the cross section at RHIC is rather low. At design luminosity for
Si+Si ($4.4 \cdot 10^{28} cm^{-2} s^{-1}$), about 300 
$\Upsilon$s are produced in a RHIC year ($10^7$ s).  The situation is
better with Pb-ions at the LHC. The design luminosity ($1 \cdot
10^{26} cm^{-2} s^{-1}$) corresponds to a production rate of about
0.02 Hz or roughly 60 $\Upsilon$s per hour.  The experimental
identification in heavy-ion interactions is relatively easy because of
the coherence requirement, which limits production to $p_T <
\sqrt{2}\hbar / R$ \cite{Adler:2002sc}.

Eq. \ref{Ryskin} shows that the forward scattering amplitude for
$\Upsilon$ production is proportional to the gluon density squared. A
30\% reduction in the nuclear gluon density would roughly halve the
cross section. Photonuclear $\Upsilon$ production at the LHC should
be a sensitive probe of nuclear gluon shadowing in the range
$x\approx 2\times10^{-3}$.

The coherent production of $\Upsilon$ at the LHC was studied recently
by Frankfurt et al. \cite{Frankfurt:2003qy}. Our result (solid curve
in Fig.~\ref{fig:AA}) is about 10\% higher than their result for the
impulse approximation (no shadowing).  The difference may be due to 
the slightly different photon spectrum and  
slope of $d \sigma/dt$ in photon-proton interactions. With nuclear
gluon shadowing, the cross section at mid-rapidity may be reduced by
as much as 50\% for Pb+Pb \cite{Frankfurt:2003qy}.

Photoproduction of other final states should also be accessible
at existing and future $\overline{p}p$ and $pp$ colliders.  For
example, photoproduction of open charm and bottom  could be
used to measure gluon distributions.  These events would have only
a single rapidity gap, but the experimental techniques should be similar. 

To summarize, we have calculated the cross sections for
photoproduction of heavy vector mesons in $p p$ and $p \overline{p}$
collisions. The cross sections are large enough for this
reaction channel to be observed experimentally. The $d \sigma / dt$ is
distinctly different in $p p$ and $p \overline{p}$ collisions because
of the interference between the production sources. The cross section
for producing $\Upsilon$ mesons in coherent photonuclear Pb+Pb
interactions at the LHC is large. 
Because of the distinctive experimental signature, these reactions
should be easy to detect.

This work was supported by the U.S. Department of Energy under
Contract No. DE-AC-03076SF00098.

\newpage 
\bibliography{apssamp}
\end{document}